# Classical Authentication Aided Three-Stage Quantum Protocol

Partha Basuchowdhuri

**Abstract:** This paper modifies Kak's three-stage protocol so that it can guarantee secure transmission of information. Although avoiding man-in-the-middle attack is our primary objective in the introduction of classical authentication inside the three-stage protocol, we also benefit from the inherent advantages of the chosen classical authentication protocol. We have tried to implement ideas like key distribution center, session key, timestamp and nonce, within the quantum cryptography protocol.

## Introduction

The purpose of cryptography is to restrict access to a message and let it be read by only the desired recipient. In classical cryptography, a message is encrypted by a key and that key is kept secret (or made public). Decrypting the message is made hard by large number of possible keys and the key is distributed among interested parties who can decrypt and get the message. Key distribution remains a difficult issue in the first place.

Quantum key distribution is a provably secure protocol, which can create a private key between parties interested to exchange information over public channel. This key can be used to create a classical private key cryptosystem. The only requirement for quantum key distribution is that, while transmitting over the public channel, the error-rate (R) must be less than a certain threshold.

In this note we modify Kak's three-stage protocol [1] so that it can guarantee secure transmission of information. We do this by the use of a hybrid model, which seems reasonable since the implementation of a quantum cryptography system will never be purely quantum. Although avoiding man-in-the-middle attack [2] is our primary objective in the introduction of classical authentication inside the three-stage protocol, we also benefit from the inherent advantages of the chosen classical authentication protocol.

## Quantum key distribution

Say an eavesdropper Eve is trying to intercept the message being transmitted between Alice and Bob. Quantum key distribution is effective because of the no-cloning theorem [3]. If Eve tries to differentiate between two non-orthogonal states, it is not possible to achieve information gain without collapsing the state of at least one of them.

This is clear from considering $|\psi\rangle$ and $|\varphi\rangle$ to be the non-orthogonal quantum states Eve is trying to know about. If these states interact with a standard state $|u\rangle$,

$$|\psi\rangle|u\rangle \rightarrow |\psi\rangle|v\rangle$$
$$|\varphi\rangle|u\rangle \rightarrow |\varphi\rangle|v'\rangle$$

Eve would want $|v\rangle$ and $|v'\rangle$ to be different, to know the identity of the state. However inner products are preserved under unitary transformations and

$$\langle v|v'\rangle\langle\psi|\varphi\rangle = \langle u|u\rangle\langle\psi|\varphi\rangle \quad \text{or,} \quad \langle v|v'\rangle = \langle u|u\rangle = 1$$

So $|v\rangle$ and $|v'\rangle$ must be identical and Eve will need to disturb one of the two states in order to acquire any information.

**Kak's three-stage quantum protocol**

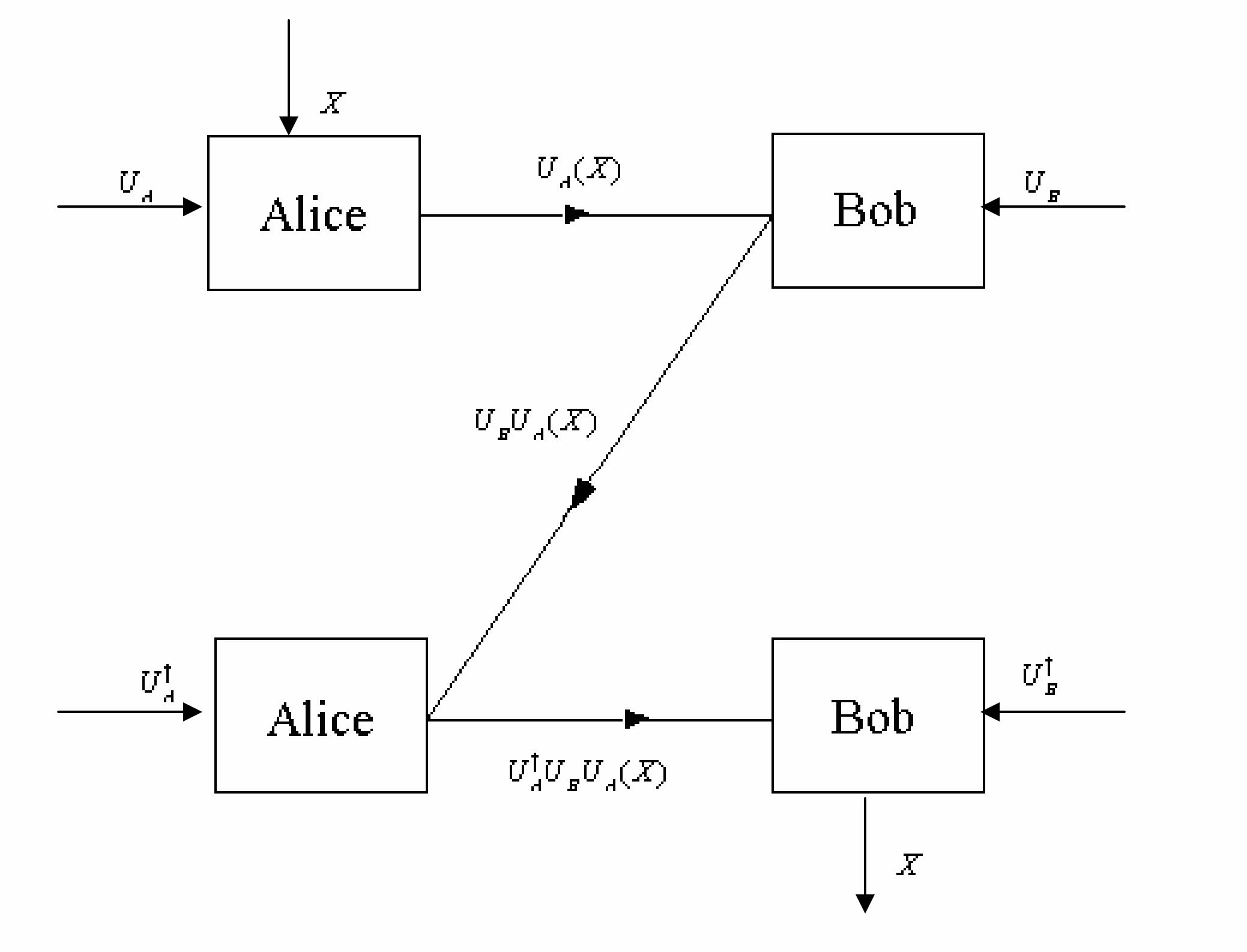

**Figure 1. Three stage quantum protocol applying commutative transformations**

This protocol can be summarized as follows:

1. Alice applies a unitary transformation $U_A$ on quantum information X and sends the qubits to Bob.
2. Bob applies $U_B$ on the received qubits $U_A$, which gives $U_B U_A(X)$ and sends it back to Alice.
3. Alice applies $U_A^\dagger$ (transpose of the complex conjugate of $U_A$) on the received qubits to get $U_A^\dagger U_B U_A(X) = U_A^\dagger U_B U_D(X) = U_B(X)$ and sends it back to Bob.
4. Bob applies $U_B^\dagger$ on $U_B(X)$ to get the information X.

Here $U_A$ and $U_B$ must be commutative to each other, which means that $U_B U_A(X) = U_A U_B(X)$.

With n number of qubits present in the message, the transformations $U_A$ and $U_B$ both must be of $2^n$ dimension. It has been observed that the (2 × 2) rotation operator, Pauli-X, Pauli-Y and Pauli-Z can be used as commutative transformations in 1-qubit system as all of these are 2 × 2 matrices.

In order to find transformations for an n qubit system we can randomly pick any of these 2×2 matrices and tensor multiply it [4] with another randomly picked one (may be itself) and keep on tensor multiplying for n times, which will eventually produce a $2^n \times 2^n$ matrix. The commutativity of the rotation operator can be shown as below:

$$R(\theta) = \begin{pmatrix} \cos\theta & -\sin\theta \\ \sin\theta & \cos\theta \end{pmatrix}$$

$$R(\theta).R(\phi) = \begin{pmatrix} \cos\theta & -\sin\theta \\ \sin\theta & \cos\theta \end{pmatrix} . \begin{pmatrix} \cos\phi & -\sin\phi \\ \sin\phi & \cos\phi \end{pmatrix} = \begin{pmatrix} \cos(\theta+\phi) & -\sin(\theta+\phi) \\ \sin(\theta+\phi) & \cos(\theta+\phi) \end{pmatrix}$$

For a 2-qubit system:

$$R_2(\theta) = \begin{pmatrix} 1 & 0 \\ 0 & 1 \end{pmatrix} * \begin{pmatrix} \cos\theta & -\sin\theta \\ \sin\theta & \cos\theta \end{pmatrix} = \begin{pmatrix} \cos\theta & -\sin\theta & 0 & 0 \\ \sin\theta & \cos\theta & 0 & 0 \\ 0 & 0 & \cos\theta & -\sin\theta \\ 0 & 0 & \sin\theta & \cos\theta \end{pmatrix}$$

where "*" denotes tensor product.

## Man-in-the-middle attack for three-stage protocol

Man-in-the-middle attack can affect both classical and quantum channels. Here Eve can pretend to be Bob to Alice and vice-versa. Instead of $U_B$ she selects $U_C$ (which is also commutative) and fakes a response which looks similar to what Bob would have done. Eve pretends as Alice to Bob with the transformation $U_D$, which is commutative to $U_B$ and instead of X sends a gibberish Y. So from interaction with Alice he acquires value X and sends a junk Y to Bob and hence disables the protocol.

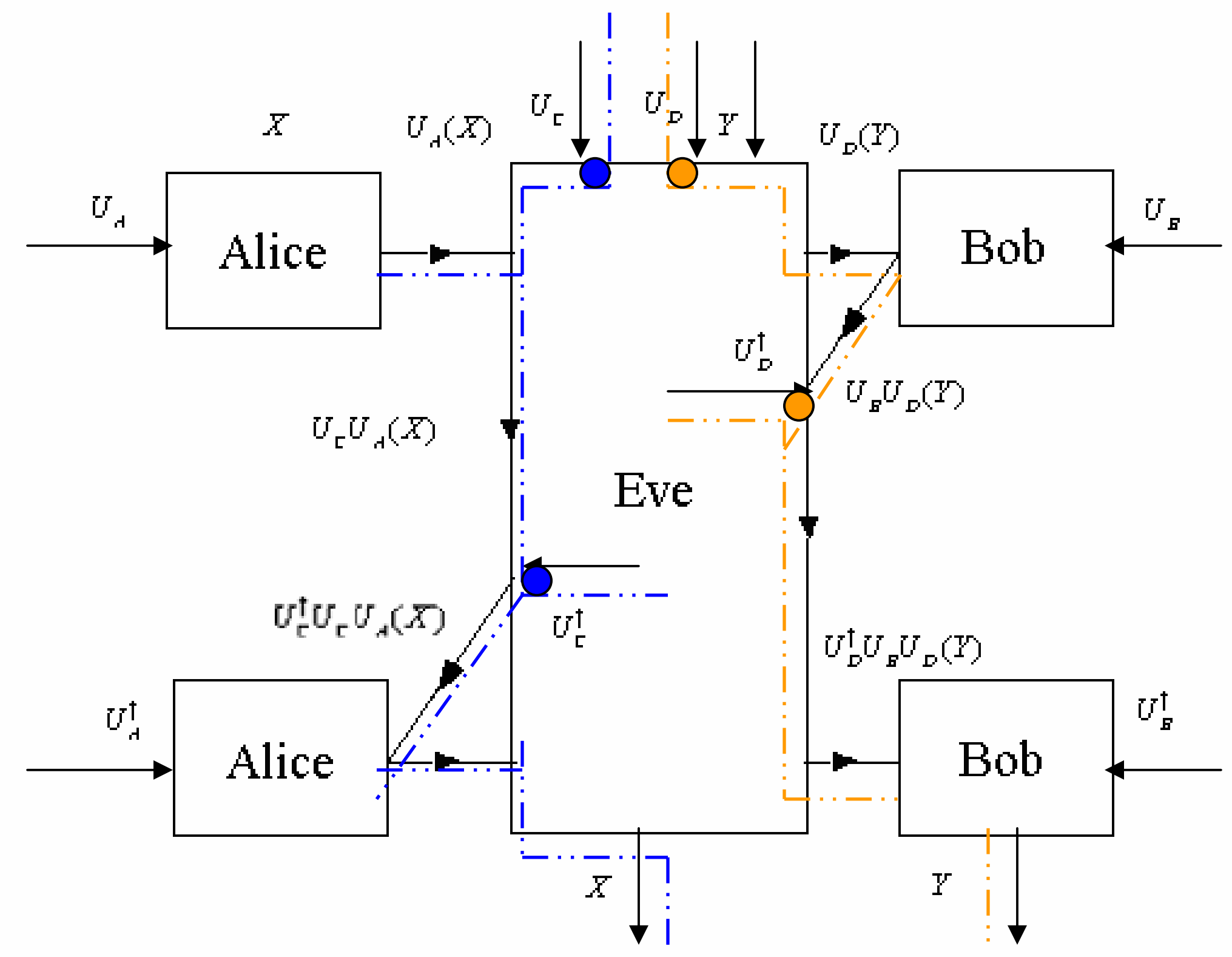

**Figure 2. Three-stage protocol under Man-in-the-middle attack**

Perkins proposed use of entangled pairs for a possibly secure three-stage protocol that can avoid man-in-the-middle attack [5]. But while distributing the entangled pairs, they might get corrupt during the transit. So if the entangled pairs are not matched then the process will be aborted and this might take place very frequently. So in order to ensure security from man-in-the-middle attack and avoid the uncertainty regarding entangled pairs this paper proposes a hybrid model that uses classical authentication protocols to ensure security in three-stage quantum protocol.

## Description of the proposed protocol

Here, we use modified version of the protocols proposed by Denning-Sacco [6] and Kehne *et al* [7], as the authentication protocol, alongside the qubits sent in each stage. It takes help from a central Key Distribution Center (KDC), which assigns the session key and work as the central authority for authentication.

Firstly, as this protocol uses classical bit sequence, we have to transform the bit sequence into qubits. A bit sequence of 01101… can be transformed into $|0\rangle|1\rangle|1\rangle|0\rangle|1\rangle...$, even; to increase the amount of reliability we can map 0 and 1 into more than one photon. Now what we are doing in each step is that we are sending a series of photon as in usual three-stage protocol and still using the authentication protocol. Each time Alice or Bob (or the KDC) gets the stream of photons; they convert the authentication part to classical information, then process it and again transform it into quantum information before transmitting. Say, Q (.) is the function used to denote conversion of classical bits to quantum qubits. $Q^{-1}(.)$ is also used to get the classical bits inside the Alice, Bob and KDC units and are not shown in the protocol.

The protocol can be described as follows:

**1.** $A \rightarrow B$: $Q(ID_A \| N_a) \| U_A(X)$

**2.** $B \rightarrow KDC$: $Q(ID_B \| N_b \| E_{K_b}[ID_A \| N_a \| T_b]) \| U_B U_A(X)$

**3.** $KDC \rightarrow A$: $Q(E_{K_a}[ID_B \| N_a \| K_s \| T_b]) \| E_{K_b}[ID_A \| K_s \| T_b]) \| N_b) \| U_B U_A(X)$

**4.** $A \rightarrow B$: $Q(E_{K_b}[ID_A \| K_s \| T_b]) \| E_{K_s}[N_b]) \| U_A^{\dagger} U_A U_B(X)$

**1.** A sends quantum information of the nonce $N_a$ and the ID along side the message to B.

**2.** B also sends his ID and nonce and encrypts A's ID, nonce and his timestamp using shared key between B and KDC. This initiates KDC to assign a session key.

**3.** KDC assigns a session key and prepares packages for A and B which include their own ID, session key and B's timestamp. A's nonce is encrypted inside A's package but B's nonce is kept open. A gets back his nonce and A is assured of its timeliness by the session key and ensured that it's not a replay. This block also verifies that B has received A's earlier message with help of B's ID.

**4.** Session key authenticates that the message came from A and is not a replay.

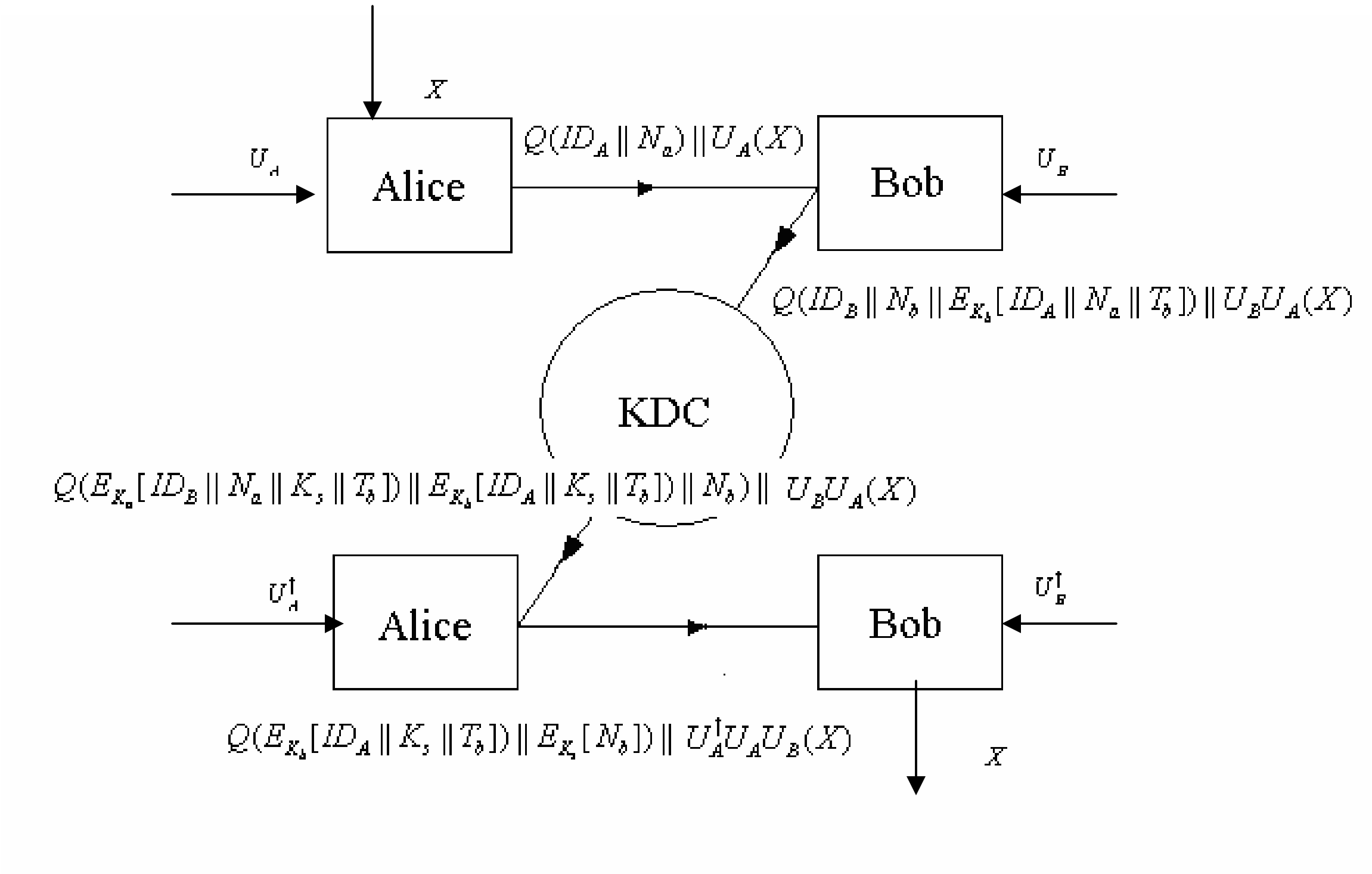

**Figure 3: Classical authentication protocol aided three-stage quantum protocol**

Under the described circumstances, man-in-the-middle attack is no more possible, because we are using timestamps, IDs, session keys, nonces and encryption key for verification. If the authentication process detects any problem, the process can be aborted. In this protocol we are dependent on the KDC and we need the KDC to be trust-worthy.

## Conclusions

There are many inherent advantages of the proposed system. For, example, the KDC can't see the message in this system. Neither Bob nor Eve is capable of modifying or forging the message. Alice cannot disavow sending that message and simultaneously Bob cannot deny receiving the message later on. The transmission is quantum and hence non-reproducible. Synchronization throughout the network is not needed as the time-stamp is provided by Bob and hence will correspond to Bob's clock only. Suppress-reply attacks can be avoided because the nonces the recipients will choose are unpredictable to the sender.

**References**

[1] S. Kak, “A three-stage quantum cryptography protocol.” Foundations of Physics Letters 19, 293, 2006; arXiv: quant-ph/0503027.

[2] K. Svozil, “Feasibility of the interlock protocol against man-in-the-middle attacks on quantum cryptography.” arXiv: quant-ph/0501062

[3] M.A Nielsen, I.L. Chuang, Quantum computation and Quantum information. Cambridge University Press, 2000.

[4] P. Sivakumar, “Implementing the three-stage quantum cryptography protocol.” arXiv: cs/0603067

[5] W. Perkins, “Trusted certificates in quantum cryptography.” arXiv: cs/0603046

[6] D. Denning and G. Sacco, “Timestamps in key distribution protocols.” Communications of the ACM, 24(8):533-536, August 1981.

[7] A. Kehne, J, Schonwalder, and H. Langendorfer, “A nonce-based protocol for multiple authentications” *Operating Systems Review*, October 1992.